\documentclass[twocolumn,showpacs,preprintnumbers,amsmath,amssymb]{revtex4-1}

\usepackage{afterpage}
\usepackage[dvipdfmx]{graphicx}
\usepackage{color,bm}

\begin{document}

\title{
Correspondence between two gravitational lens equations in
a static and spherically symmetric spacetime
}
\author{Ryuya Kudo}
\email{kudo@tap.st.hirosaki-u.ac.jp}
\author{Hideki Asada}
\email{asada@hirosaki-u.ac.jp}
\affiliation{
Graduate School of Science and Technology, Hirosaki University,
Aomori 036-8561, Japan}
\date{\today}

\begin{abstract}
Virbhadra and Ellis have proposed a very accurate equation
(referred to as VE equation)
for the gravitational lens in a static and spherically symmetric spacetime
[Phys. Rev. D 62,084003 (2000)],
whereas an improved equation (referred to as
OB equation) has been derived by Bozza
[Phys. Rev. D 78, 103005 (2008)]
based on a relation found by Ohanian
[Am. J. Phys. 55, 428 (1987)].
OB equation was rediscovered later
by Takizawa, Ono and Asada
[Phys. Rev. D, 102, 064060 (2020)].
VE and
OB equations seem to be very different from each other.
The present paper shows that there exists an unphysical branch
in VE equation.
Consequently, VE equation can be
improved by removing the unphysical branch.
The improved version of VE equation is found to be the same as
OB equation
when a suitable transformation is made between the deflection angles
defined differently in the two formulations.
An explicit expression of the transformation is found.
We also argue possible numerical
differences when the transformation between the deflection angles is ignored.
\end{abstract}


\maketitle

\section{Introduction}
The gravitational lens plays crucial roles in modern cosmology
and gravitational physics
\cite{SEF,Petters,Dodelson,Keeton,Will},
where most of earlier works focus on weak deflection of light.
However, the direct imaging of the massive black hole candidates
by the Event Horizon Telescope has drastically changed the situation
\cite{EHT2019,EHT2022}.
Exact (or very accurate) treatments are needed not only for theoretical studies
\cite{Darwin,Bozza,Perlick}
but also for observational researches on the gravitational lens
in the strong gravity region.
Notably, the project of the next generation Event Horizon Telescope is
increasing the importance of the lens by strong gravity
\cite{ngEHT}.

As a pioneering work,
Virbhadra and Ellis (VE)  have proposed a very accurate equation
for the gravitational lens in a static and spherically symmetric spacetime
\cite{VE}.
Their geometrical approach is almost exact
\cite{Perlick},
though they assume that the deflection of light is caused
on the lens plane.
Their assumption is a good approximation
when the observer, lens and source configuration is almost aligned.
Recently, VE equation has been used for
detailed discussions of image distortions
of Schwarzschild lensing
\cite{Virbhadra2022, Virbhadra2024}.

Instead of assuming such a symmetrical configuration,
Bozza has made use of the Ohanian relation
for a more general configuration
\cite{Ohanian}
to obtain an improved version of the accurate lens equation
\cite{Bozza2008}.
For its simplicity,
the improved equation is referred to as
Ohanian-Bozza (OB) equation.
Takizawa, Ono and Asada
have later rediscovered this equation
in an apparently different form
by using a state-of-art formulation
\cite{GW,Ishihara2016,Ishihara2017} based on the Gauss-Bonnet theorem
\cite{Takizawa2020}.
OB equation seems to be very different from VE equation.
Is it true?

The main purpose of this paper is
to examine whether VE equation
is really different from OB one.
First, we shall show that
VE equation has a duality that it has not only a physical branch
but also an unphysical one.
Hence, VE equation can be improved
by removing the unphysical branch.
Secondly,
we shall argue that
the improved version of VE equation can be the same as OB equation
when we make a suitable transformation between the deflection angles defined differently
in the two formulations.

This paper is organized as follows.
Section II discusses the unphysical branch in VE equation.
Section III clarifies a relation between the improved VE equation
and OB one.
Section IV argues
possible numerical differences
when the deflection angle transformation is ignored.
Appendix A shows that there
exists no solution corresponding to a lensed image
in the unphysical branch of VE equation.
Throughout this paper,
we focus on a static and spherically symmetric spacetime.

\section{Two lens equations}
\subsection{VE and OB equations}
\begin{figure}
\includegraphics[width=8.5cm]{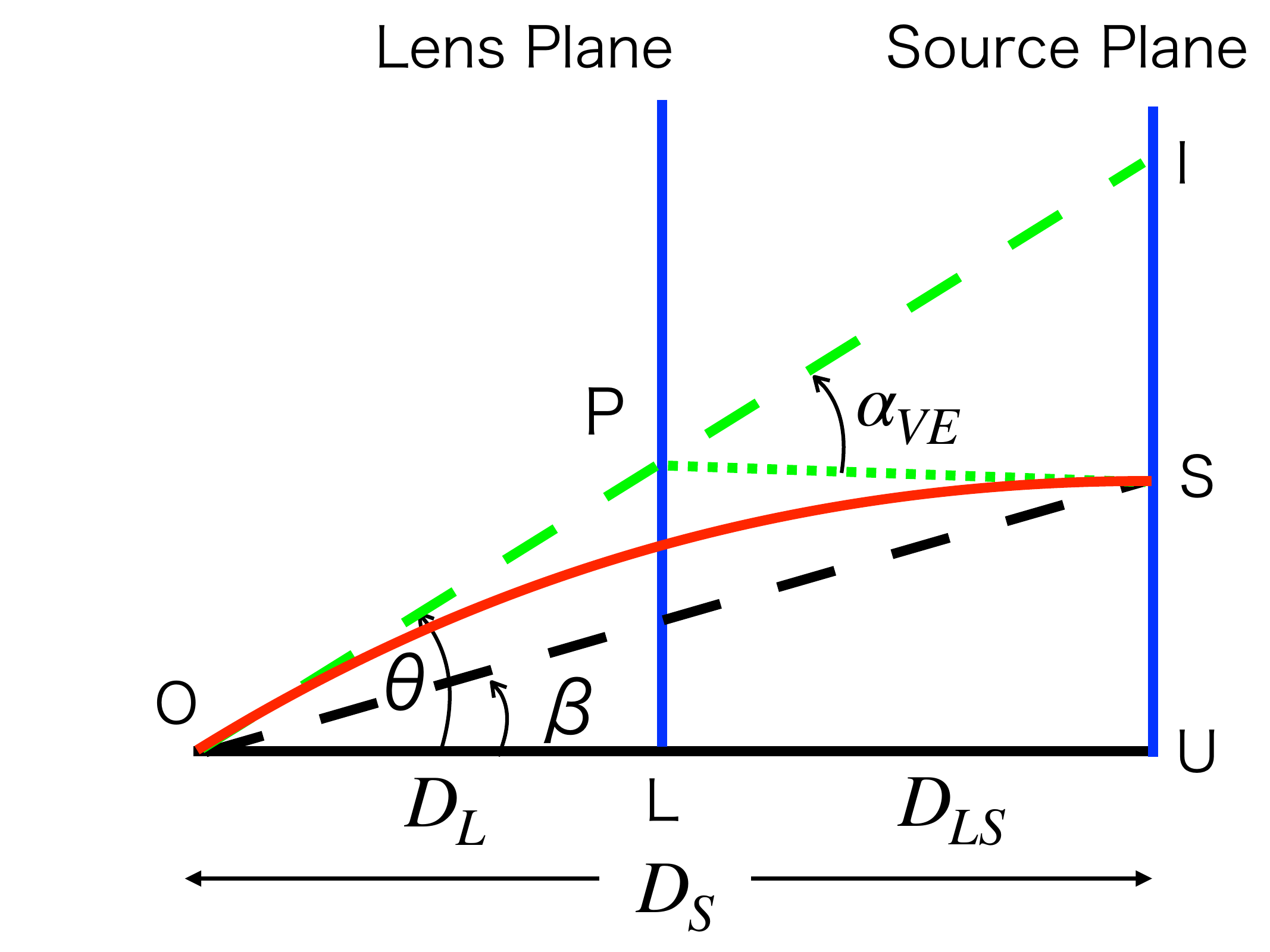}
\caption{
The deflection occurs at the point P on the lens plane.
The deflection angle $\alpha_{VE}$ is assumed to be positive
in this figure.
}
\label{fig-VE}
\end{figure}

Let us begin with briefly summarizing
VE and OB equations.
Both equations describe the mapping between
the source angle $\beta$ and image angle $\theta$
in terms of angular diameter distances
in rigorous manners
(e.g. See Figure \ref{fig-VE}).
$D_L$, $D_S$ and $D_{LS}$ denote
the angular diameter distances
between the observer and the lens plane,
between the observer and the source plane,
and
between the lens and the source plane, respectively.

A crucial difference between the two equations is
how to treat the deflection angles of light.
See Figure \ref{fig-VE}.
VE equation assumes
an almost aligned configuration in which the observer and source are
equidistant from the lens,
as usually assumed in the conventional lens equation
with the thin lens approximation \cite{SEF}.

In Figure \ref{fig-VE},
we consider a relation among the lengths $IS$, $SU$ and $IU$
on the source plane as $IS = IU - SU$.
From this relation, VE equation can be obtained as
\begin{align}
D_S \tan\beta
=
D_S \tan\theta - D_{LS}
[\tan\theta + \tan(\alpha_{VE} - \theta)] ,
\label{lenseq-VE}
\end{align}
where $\alpha_{VE}$ is an angle measured at the point P on the lens plane
in Figure \ref{fig-VE}.
From the aspect of the triangular inequality,
Dabrowski and Schunck pointed out difficulties of using VE equation
\cite{DS}.

\begin{figure}
\includegraphics[width=8.5cm]{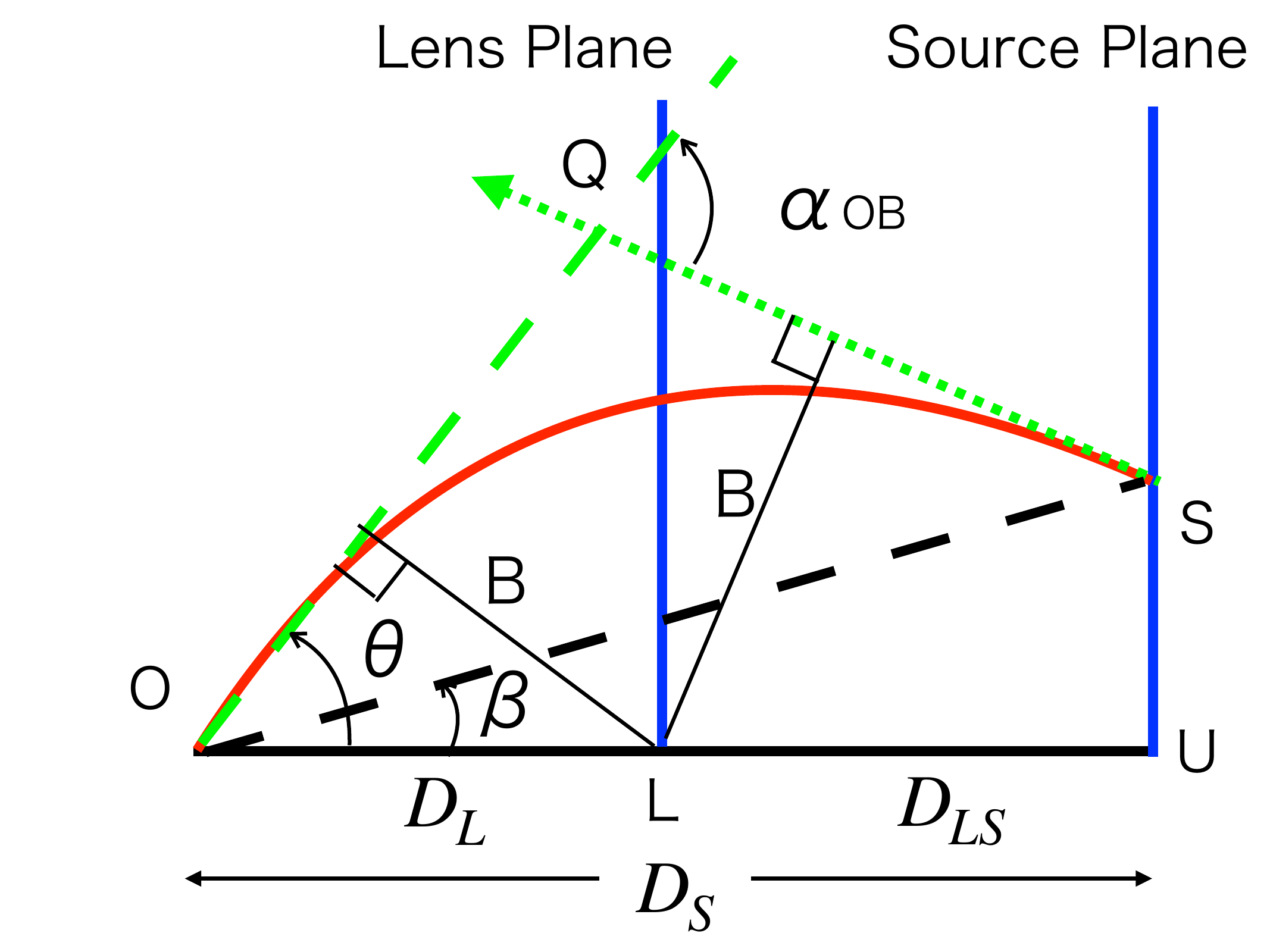}
\caption{
The deflection point Q in the OB equation
is off the lens plane.
}
\label{fig-Bozza}
\end{figure}

In order to avoid an almost aligned assumption,
Ohanian considered the tangent line at the source and that at the observer
(See Figure \ref{fig-Bozza}) and the intersection point Q
\cite{Ohanian}.
Note that the intersection point Q is generally off the lens plane.
For the quadrilateral OLSQ (See Figure \ref{fig-Bozza}),
a relation among the four interior angles can be obtained.
It is Ohanian relation \cite{Ohanian}.
Bozza rewrote it in terms of the angular diameter distances
to arrive at an improved version of an accurate lens equation as
\begin{align}
D_S \tan\beta
=
\frac{D_L \sin \theta - D_{LS} \sin(\alpha_{OB} - \theta)}{\cos(\alpha_{OB} - \theta)} ,
\label{lenseq-Bozza}
\end{align}
where $\alpha_{OB}$ is defined at the intersection point Q
in Figure \ref{fig-Bozza}.

By using the Gauss-Bonnet method
on the gravitational lens \cite{GW},
another accurate lens equation is derived as
\cite{Takizawa2020}
\begin{align}
&\alpha_{TOA} - \theta
- \arcsin\left(\frac{D_L}{\sqrt{(D_{LS})^2 + (D_S)^2 \tan^2 \beta}} \sin \theta \right)
\notag\\
&+ \arctan \left( \frac{D_S}{D_{LS}} \tan\beta \right)
\notag\\
&= 0 ,
\label{lenseq-TOA}
\end{align}
where
$\alpha_{TOA}$ is defined at the intersection point Q
in Figure \ref{fig-Bozza}.
This equation is equivalent to Eq. (\ref{lenseq-Bozza}).
See Reference \cite{Takizawa2020}
for a proof of the equivalence.
Hence, $\alpha_{OB} = \alpha_{TOA}$.

\subsection{Duality in VE equation}
By straightforward calculations,
Eq. (\ref{lenseq-VE}) is rewritten into
a quadratic in $\tan\theta$ as
\begin{align}
&D_L \tan\alpha_{VE} (\tan\theta)^2
+ D_S (1 - \tan\beta \tan\alpha_{VE}) \tan\theta
\notag\\
&- (D_S \tan\beta + D_{LS} \tan\alpha_{VE}) = 0 .
\label{lenseq-VE-square}
\end{align}
In Figure \ref{fig-VE},
we may expect that a positive $\tan\theta$ existed uniquely
for a given $\alpha_{VE}$.
However, it is not the case.
Eq. (\ref{lenseq-VE}) is equivalent to Eq. (\ref{lenseq-VE-square})
and it thus admits two $\tan\theta$.
One of the two $\tan\theta$ corresponds to Figure \ref{fig-VE},
whereas the other $\tan\theta$ is unphysical
because it does not describe the light ray in Figure \ref{fig-VE}.

For $\alpha_{VE}$ in Figure \ref{fig-VE},
Eq. (\ref{lenseq-VE-square}) is formally solved for $\tan\theta$ as
\begin{align}
\tan\theta
=
\frac{D_S (\tan\beta \tan\alpha_{VE} - 1) \pm \sqrt{\cal D}}{2 D_L \tan\alpha_{VE}} ,
\label{lenseq-VE-linear}
\end{align}
where
\begin{align}
{\cal D}
\equiv &
(D_S)^2 (\tan\beta \tan\alpha_{VE} - 1)^2
\notag\\
&+ 4 D_L \tan\alpha_{VE} (D_S \tan\beta + D_{LS} \tan\alpha_{VE}) .
\label{D}
\end{align}
By direct calculations,
${\cal D}$ can be rewritten as
\begin{align}
{\cal D}
=
A \left(\tan\alpha_{VE} - \frac{B}{A}\right)^2
+ \frac{4D_L D_{LS}  (D_S)^2}{A \cos^2\beta} ,
\label{D2}
\end{align}
where $A \equiv (D_S)^2 \tan^2\beta + 4 D_L D_{LS}$,
$B \equiv  D_S (D_{LS} - D_L) \tan\beta$, and $A>0$.
Eq. (\ref{D2}) gives ${\cal D} > 0$
for any $\alpha_{VE}$.

The plus and minus signs in Eq. (\ref{lenseq-VE-linear})
reflect two branches in VE equation.
Note that Eq. (\ref{lenseq-VE-linear}) does not mean a solution for the lens equation
for which $\alpha_{VE}$ is not a given number but it should be treated as a function of $\theta$.
Eq. (\ref{lenseq-VE-linear}) is still an equation for $\theta$ as shown later.

\subsection{Unphysical configuration for VE equation}
\begin{figure}
\includegraphics[width=8.5cm]{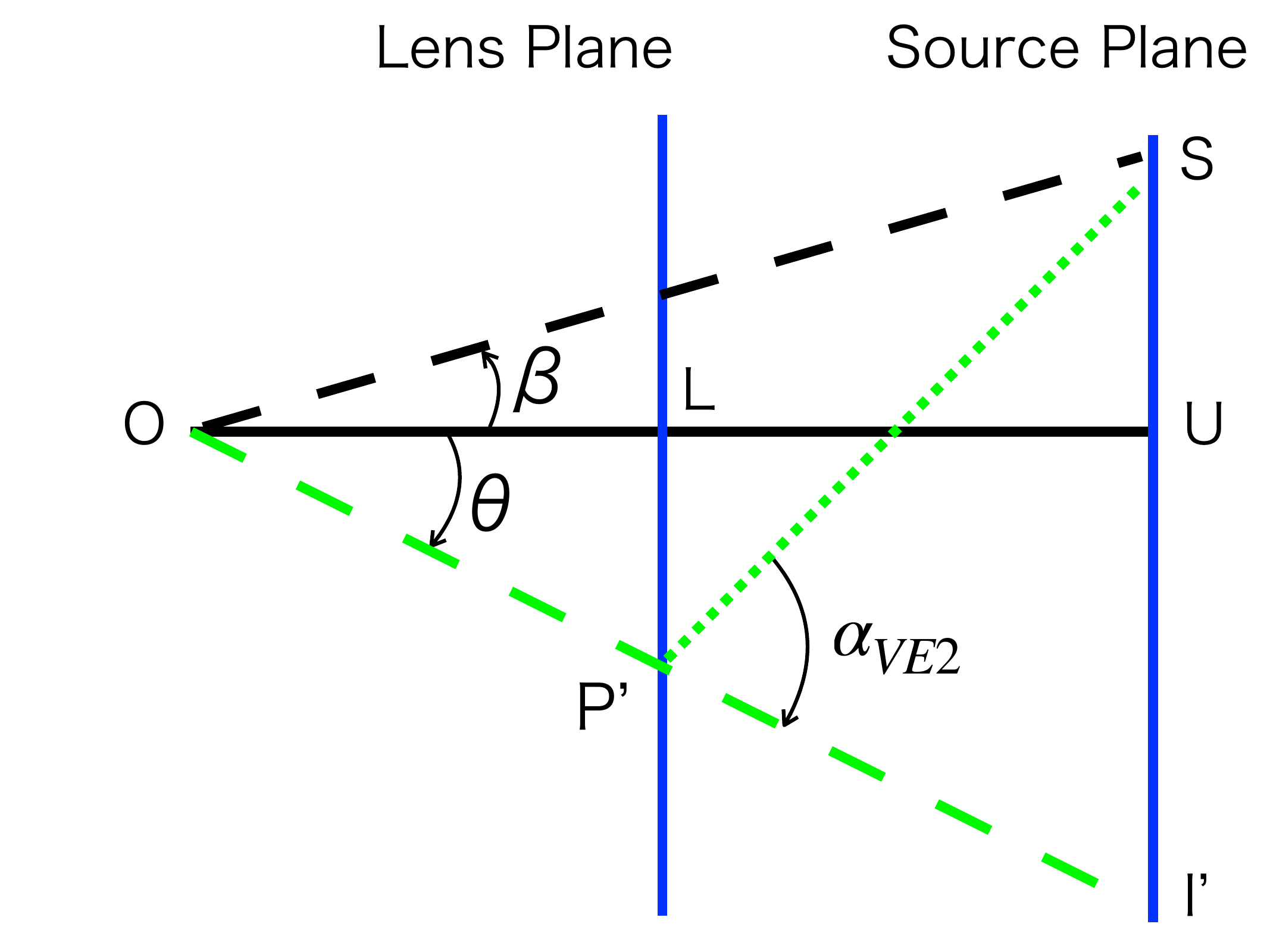}
\caption{
Another deflection point $P'$
is located on the lens plane.
Here, $\alpha_{VE2}$ is negative.
}
\label{fig-VE2}
\end{figure}

In this subsection,
we shall examine what is the unphysical branch in VE equation.
First, we consider another configuration of the observer, lens and source system
shown by Figure \ref{fig-VE2},
where $\alpha_{VE2}$ denotes the deflection angle.
We note that $0 < \alpha_{VE} <\pi$ and $-\pi < \alpha_{VE2} < 0$
in Figures \ref{fig-VE} and \ref{fig-VE2}, respectively.
For the strong deflection of a relativistic image,
$2N \pi < \alpha_{VE} < (2N+1)\pi$
in Figure \ref{fig-VE}
and
$- (2N+1) \pi < \alpha_{VE2} < - 2N \pi$
in Figure \ref{fig-VE2}
for $N =1, 2, \cdots$,
where $N$ denotes the winding number of the light ray.
We consider the deflection of light, for which
$\alpha_{VE} \neq 0$ and $\alpha_{VE2} \neq 0$.

Here, let us set $\alpha_{VE2} = \alpha_{VE} - \pi$
($\alpha_{VE2} = \alpha_{VE} - (4N+1) \pi$
for a pair of relativistic images with the winding number $N$),
which leads to $\tan\alpha_{VE2} = \tan\alpha_{VE}$.
This means that the different configuration
specified by $\alpha_{VE2} (\neq \alpha_{VE})$
satisfies the same lens equation as Eq. (\ref{lenseq-VE}).
Therefore, Eq. (\ref{lenseq-VE}) for $\tan\theta$
has two solutions,
where one solution corresponds to Figure \ref{fig-VE}
and the other describes Figure \ref{fig-VE2}.
The angle $\theta$ for the physical configuration in Figure \ref{fig-VE}
is greater than $\theta$ for the unphysical configuration in Figure \ref{fig-VE2}.

In the weak deflection of light,
the image in Figure \ref{fig-VE} is called the primary image
on the same side of the intrinsic source position
with respect to the lens
\cite{SEF}.
Hence, $\tan\theta$ in  Figure \ref{fig-VE} is greater than
that in Figure \ref{fig-VE2}.
Up to this point, we have focused on $\alpha_{VE} > 0$.

Next, we consider the case of $\alpha_{VE} < 0$,
for which Figure \ref{fig-VE2} corresponds to
a physical configuration,
whereas Figure \ref{fig-VE} is unphysical.
In the weak deflection, therefore,
the image in Figure \ref{fig-VE2} is the secondary image
on the opposite side of the intrinsic source position
with respect to the lens
\cite{SEF}.

\subsection{Duality in VE solutions}
Next, let us classify two branches
in VE equation.
For our graphical method,
it is convenient to rewrite Eq. (\ref{lenseq-VE}) into
\begin{align}
\frac{D_L}{D_{LS}} \tan\theta - \frac{D_S}{D_{LS}}  \tan\beta
=
\frac{\tan\alpha_{VE} - \tan\theta}{1 + \tan\alpha_{VE} \tan\theta} ,
\label{lenseq-VE2}
\end{align}
where $D_S - D_{LS} = D_L$ is used.

It is worthwhile to note that
$\tan\theta$ satisfying Eq. (\ref{lenseq-VE2})
corresponds to an intersection point between a straight line as
\begin{align}
y = a x - b ,
\label{y1}
\end{align}
and a hyperbolic curve as
\begin{align}
y = \frac{c -x}{1 + c x} ,
\label{y2}
\end{align}
where
$x \equiv \tan\theta$,
$a \equiv D_L/D_{LS} \geq 0$,
$b \equiv D_S \tan\beta /D_{LS}$,
and $c \equiv \tan\alpha_{VE}$.

In the rest of the paper, we assume
$\tan\beta \geq 0$ without loss of generality,
because the  spacetime under study is spherically symmetric.
Hence, $b \geq 0$.
On the other hand, the sign of $c$ is arbitrary,
because $-\infty < \tan\alpha_{VE} < + \infty$.
$c = \tan\alpha_{VE} \neq 0$ because the light is deflected.
Hence, $c >0$ or $c < 0$.

\begin{figure}
\includegraphics[width=8.5cm]{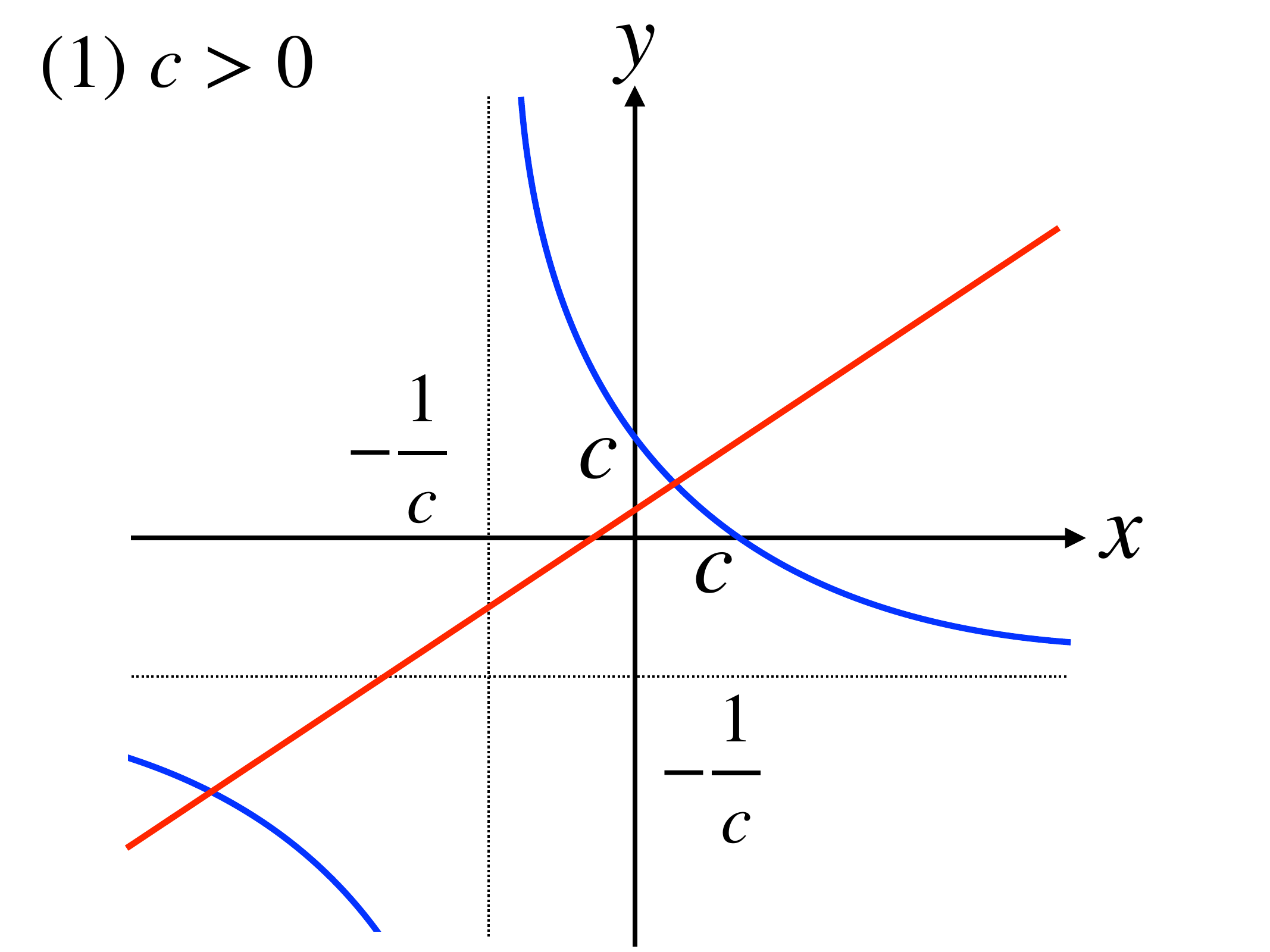}
\caption{
$c \equiv \tan\alpha_{VE} > 0$.
There are two solutions of VE equation .
The red (in color) line denotes Eq. (\ref{y1})
and the blue (in color) curves mean Eq. (\ref{y2}).
They have two intersection points.
One of the intersections is $x > 0$
and the other is $x < 0$.
}
\label{fig-curves1}
\end{figure}

There are four possible types of intersections: \\
\noindent
(1) For $c > 0$,
see Figure \ref{fig-curves1} for schematic plots of Eqs. (\ref{y1}) and (\ref{y2}).
This figure shows that
$x (= \tan\theta)$ of one intersection is positive,
whereas the other is negative.
The former and latter intersections correspond to
Figures \ref{fig-VE} and \ref{fig-VE2}, respectively.
\\

\begin{figure}
\includegraphics[width=8.5cm]{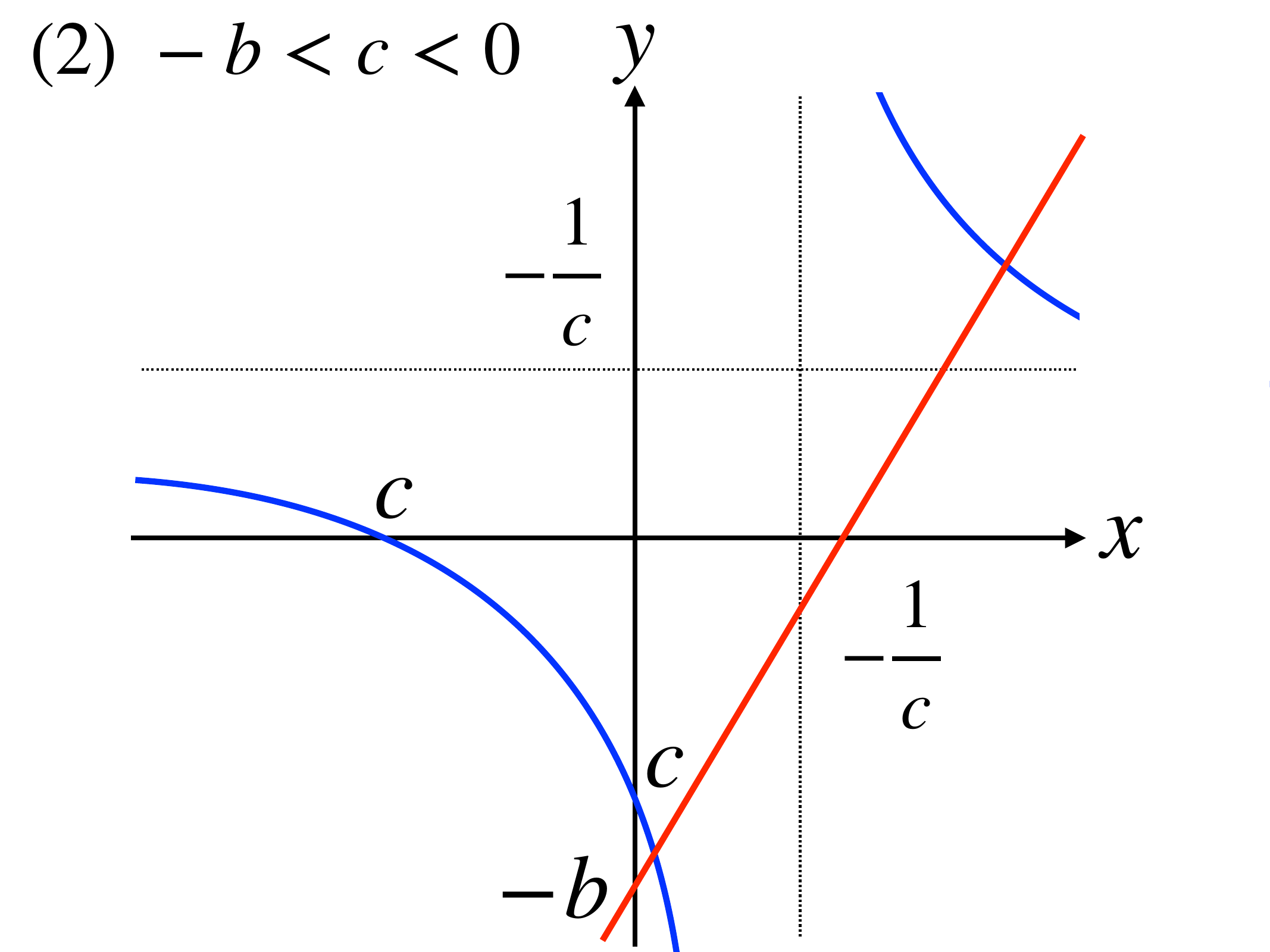}
\caption{
$-b  < c < 0$,
where $b = D_S (\tan\beta) /D_{LS}$.
There are two solutions of VE equation.
The red (in color) line denotes Eq. (\ref{y1})
and the blue (in color) curves mean Eq. (\ref{y2}).
They have two intersection points.
Both of the intersections are $x > 0$.
}
\label{fig-curves2}
\end{figure}

\noindent
(2) For $-b < c < 0$,
see Figure \ref{fig-curves2}.
$x$ for both solutions is positive.
Only for this case, there are no solutions corresponding to
Figure \ref{fig-VE2}. \\

\noindent
(3) For $-b = c < 0$,
one solution for $x$ is positive
and the other is $x=0$.
However, the latter case is corresponding to
the lens direction $\theta = 0$.
However, this direction is not observable and thus can be ignored. \\

\begin{figure}
\includegraphics[width=8.5cm]{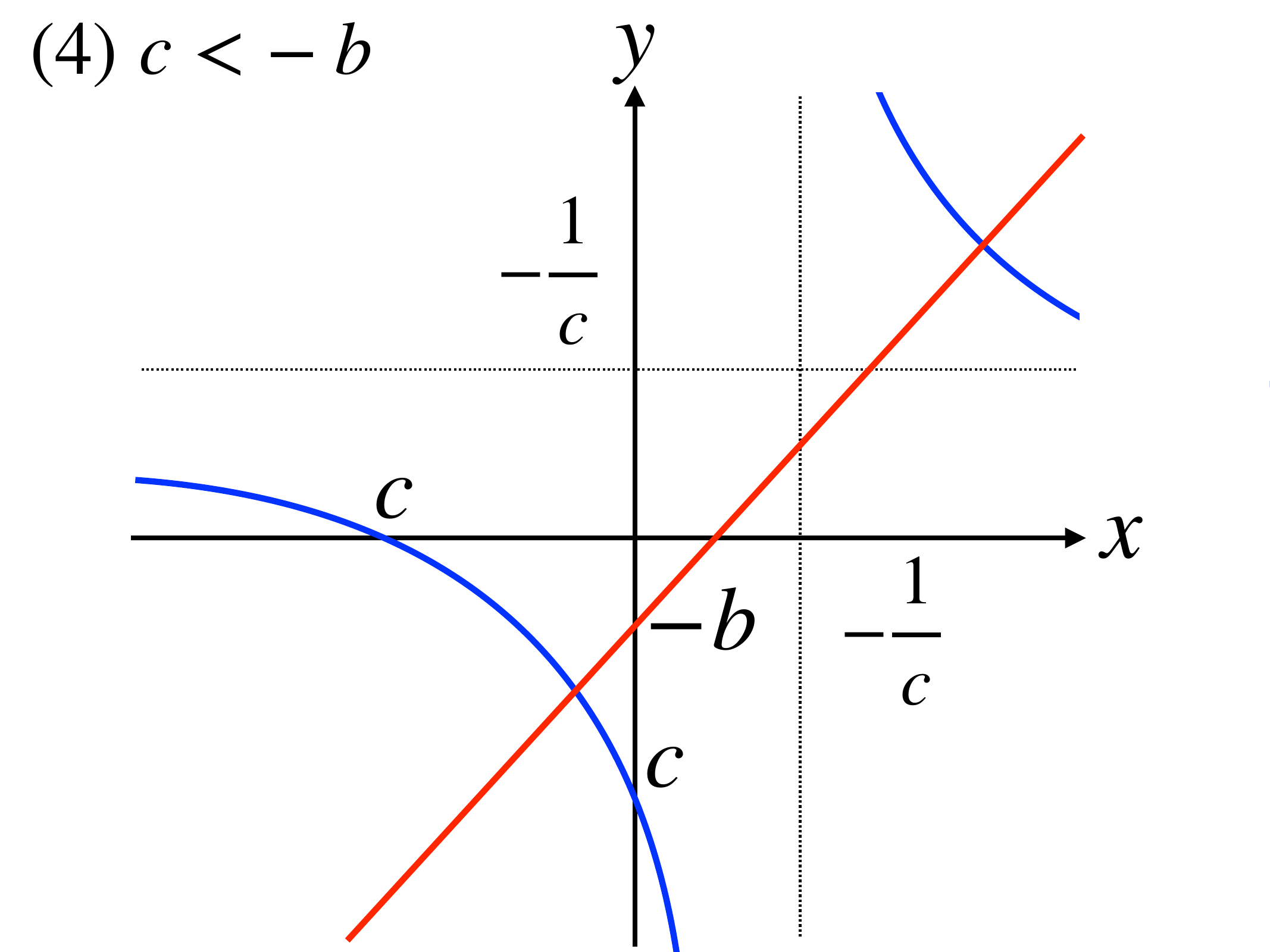}
\caption{
$c < -b$. There are two solutions of VE equation.
The red (in color) line denotes Eq. (\ref{y1})
and the blue (in color) curves mean Eq. (\ref{y2}).
They have two intersection points.
One of the intersections is $x > 0$
and the other is $x < 0$.
}
\label{fig-curves4}
\end{figure}

\noindent
(4) For $c <  -b$,
see Figure \ref{fig-curves4}.
One solution is positive and the other is negative.

In schematic plots in Figures \ref{fig-curves1}-\ref{fig-curves4},
$\tan\theta$ at an intersection is a real number.

Note that all of the above (1)-(4) are not necessarily realized.
Care is needed for the case (2) as follows.
In Figure \ref{fig-VE},
we may expect that a positive $\tan\theta$ existed uniquely
for given $\alpha_{VE}$.
However, the number of positive $\tan\theta$ in the case (2) is two.
This means that one $\tan\theta$ corresponds to Figure  \ref{fig-VE},
while the other $\tan\theta$ is unphysical
because it does not correspond to Figure \ref{fig-VE}.

In deriving Eq. (\ref{lenseq-VE}), moreover,
Figure \ref{fig-VE} is assumed,
where $\tan\theta \geq 0$ is implicitly assumed.
For this setup, therefore,
the negative $x$ in the above cases (1) and (4)
must be excluded.
Therefore,
there can exist an unphysical branch in Eq. (\ref{lenseq-VE}).

We mention the other case, $\tan\theta < 0$.
This case cannot be described by Eq. (\ref{lenseq-VE})
when $\alpha_{VE}$ is the same as that for $\tan\theta \geq 0$.
Namely,  Figure \ref{fig-VE} should be replaced by Figure \ref{fig-VE2},
which leads to another VE equation with
another $\alpha_{VE2}$
indicated in Figure \ref{fig-VE2}.
$\alpha_{VE2}$ does not equal to $\alpha_{VE}$.

\subsection{Improved VE equation}
There are plus and minus signs in Eq. (\ref{lenseq-VE-linear}).
Which sign should be chosen?
The answer relies upon the sign of $\alpha_{VE}$.
We consider $\alpha_{VE} \neq 0$ because we discuss the nonzero deflection of light.

\noindent
(1)$\alpha_{VE} >0$: \\
This case corresponds to Figure \ref{fig-VE}.
The plus sign in Eq. (\ref{lenseq-VE-linear}) should be thus chosen as
\begin{align}
\tan\theta
=
\frac{D_S (\tan\beta \tan\alpha_{VE} - 1) + \sqrt{\cal D}}{2 D_L \tan\alpha_{VE}} .
\label{lenseq-VE-plus}
\end{align}

\noindent
(2)$\alpha_{VE}<0$: \\
This case means Figure \ref{fig-VE2}.
The minus sign in Eq. (\ref{lenseq-VE-linear}) should be thus chosen as
\begin{align}
\tan\theta
=
\frac{D_S (\tan\beta \tan\alpha_{VE} - 1) - \sqrt{\cal D}}{2 D_L \tan\alpha_{VE}} .
\label{lenseq-VE-minus}
\end{align}

\section{Correspondence between VE and OB equations}
\subsection{Transformation between the VE and OB deflection angles}
\begin{figure}
\includegraphics[width=8.5cm]{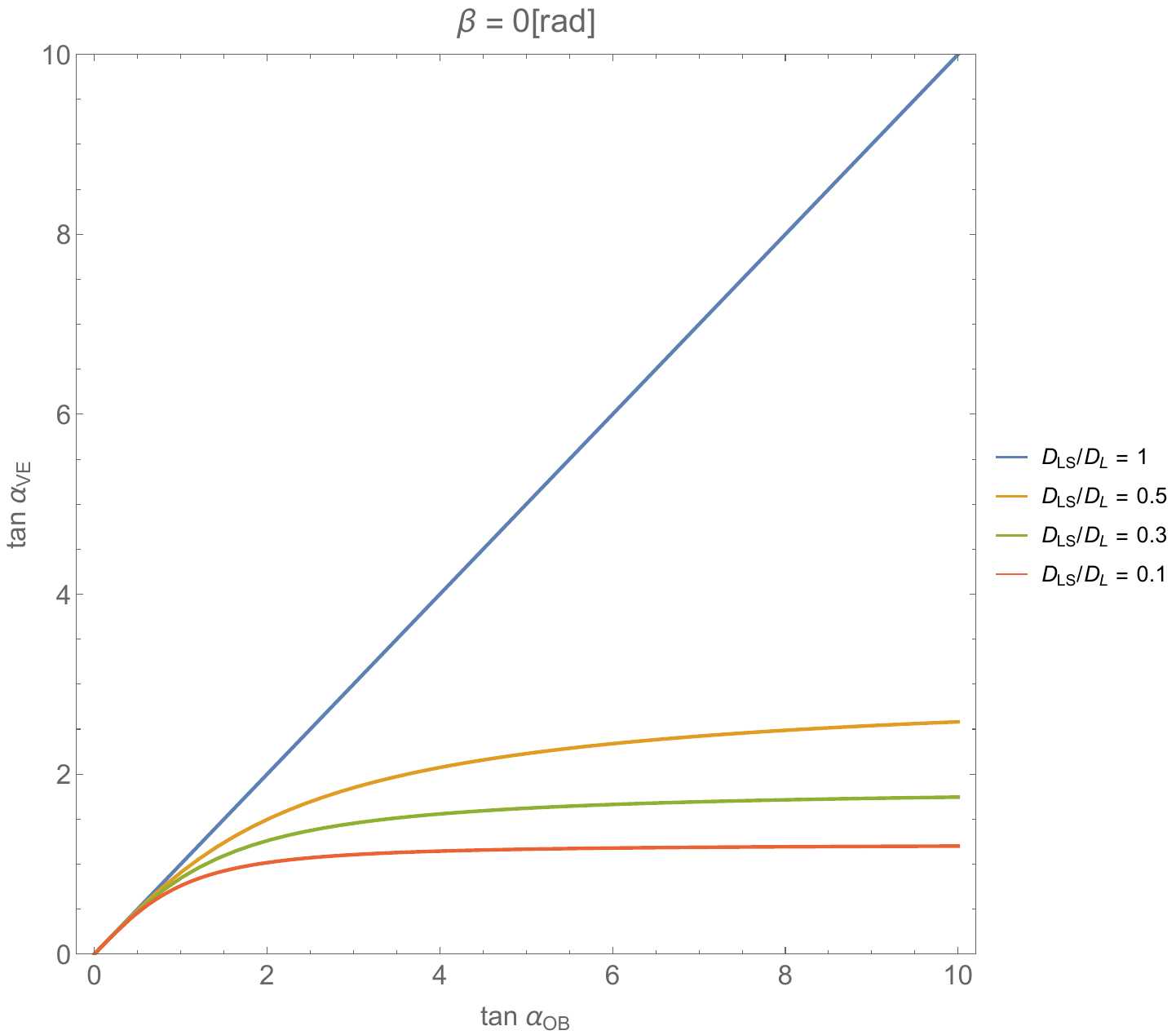}
\includegraphics[width=8.5cm]{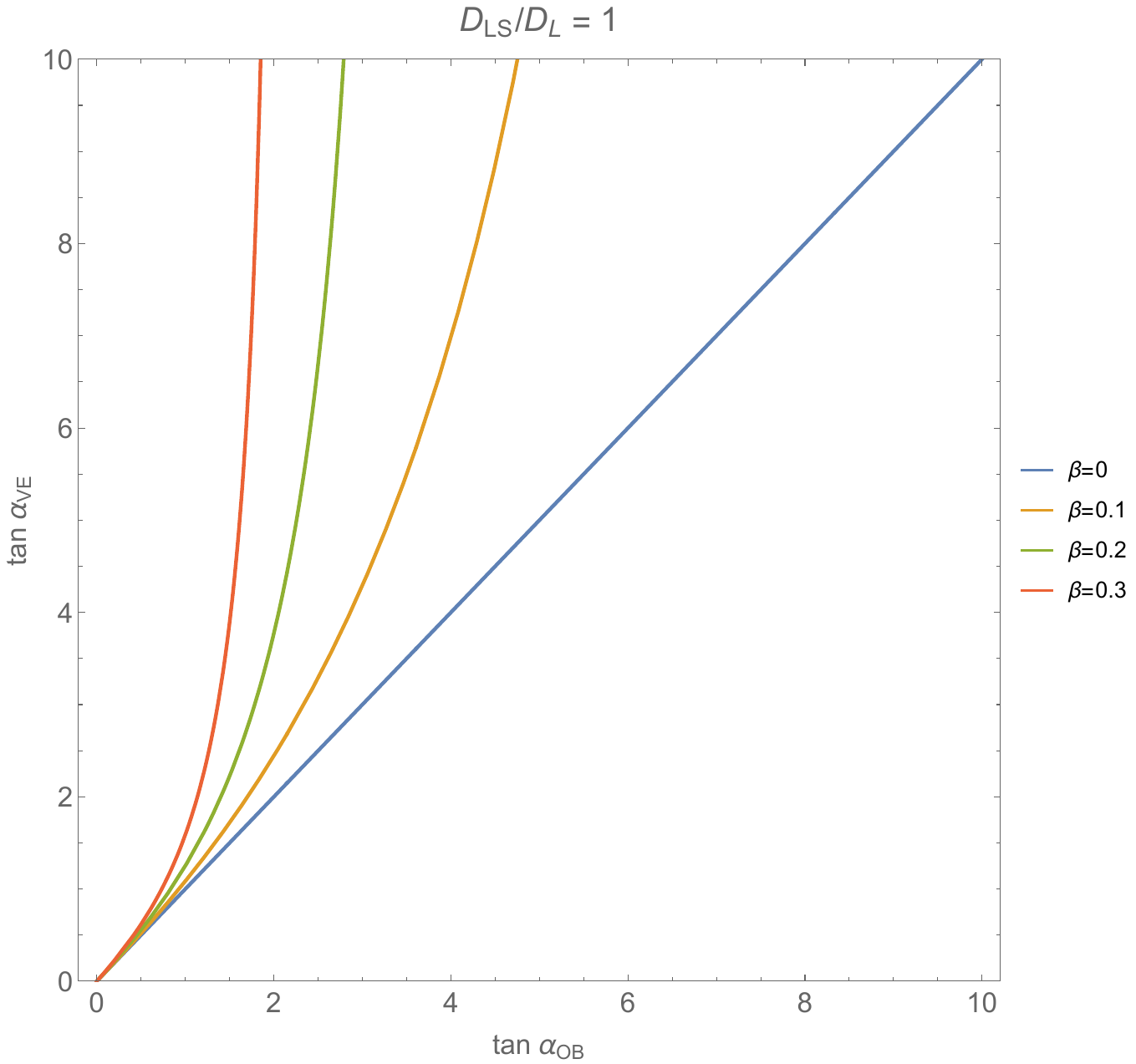}
\caption{
$\tan\alpha_{VE}$ vs $\tan\alpha_{OB}$.
Top:
$\beta =0$, and $D_{LS}/D_L =$
1 (blue), 0.5 (orange), 0.3 (green) and 0.1 (red).
Bottom:
The equidistance case.
$\beta =$ 0 (blue), 0.1 (orange), 0.2 (green) and 0.3 (red).
}
\label{fig-alpha}
\end{figure}

For the same configuration of the lens, observer and source
with given $\beta$ and $\theta$,
the improved VE and OB equations should equal to each other.
For Eq. (\ref{lenseq-VE-plus}) to equal to Eq. (\ref{lenseq-Bozza}),
a relation between $\alpha_{VE}$ and $\alpha_{OB}$ must be
\begin{align}
\tan\alpha_{VE}
= \frac{D_S F (E - F \tan\beta)}{D_{LS} F^2 + D_S E F \tan\beta - D_L E^2} ,
\label{transformation}
\end{align}
where functions $E$ and $F$ of $\alpha_{OB}$ are defined as
\begin{align}
E
&\equiv
D_{LS} \sin\alpha_{OB} + D_S \tan\beta \cos\alpha_{OB} ,
\label{E}
\\
F
&\equiv
D_L + D_{LS} \cos\alpha_{OB} - D_S \tan\beta \sin\alpha_{OB} .
\label{F}
\end{align}
Figure \ref{fig-alpha} shows a relation between
$\alpha_{VE}$ and $\alpha_{OB}$
in Eq. (\ref{transformation}).

It is worthwhile to mention that
OB equation as Eq. (\ref{lenseq-Bozza})
can be simply rewritten in terms of $E$ and $F$ as
\begin{align}
\tan\theta = \frac{E}{F} .
\label{lenseq-Bozza2}
\end{align}

\section{Numerical differences by ignoring the deflection angle transformation}
First, we consider two cases
$D_L \ll D_{LS}$ and $D_L \gg D_{LS}$.
For $D_L \ll D_{LS}$,
Eq. (\ref{transformation}) is expanded in terms of $D_L/D_{LS}$ as
\begin{align}
\tan\alpha_{VE} - \tan\alpha_{OB}
= O\left(\frac{D_L}{D_{LS}}\right) .
\label{exp1}
\end{align}
Hence, $\alpha_{VE}$ is very close to $\alpha_{OB}$.
This means that
the substitution of $\alpha_{OB}$ into $\alpha_{VE}$
in Eq. (\ref{lenseq-VE})
can be a good approximation
even if the transformation is neglected.

For $D_L \gg D_{LS}$,
on the other hand,
Eq. (\ref{transformation}) is expanded in terms of $D_{LS}/D_L$ as
\begin{align}
\tan\alpha_{VE} - \tan\alpha_{OB}
=
-\frac{1}{\tan\beta\cos\alpha_{OB}}
+ O\left(\frac{D_{LS}}{D_L}\right) .
\label{exp2}
\end{align}
This implies that
the difference between $\alpha_{VE}$ and $\alpha_{OB}$
can be significantly large
in the limit as $D_{LS}/D_L \to 0$,
especially for nearly coaligned configurations $\beta \sim 0$.
This case of $D_L \gg D_{LS}$ and $\beta \sim 0$
are corresponding to
the observation of photons from centers of galaxies,
notably a vicinity of a photon surface of
the massive black hole candidate in Sgr A${}^*$.

\begin{figure}
\includegraphics[width=8.5cm]{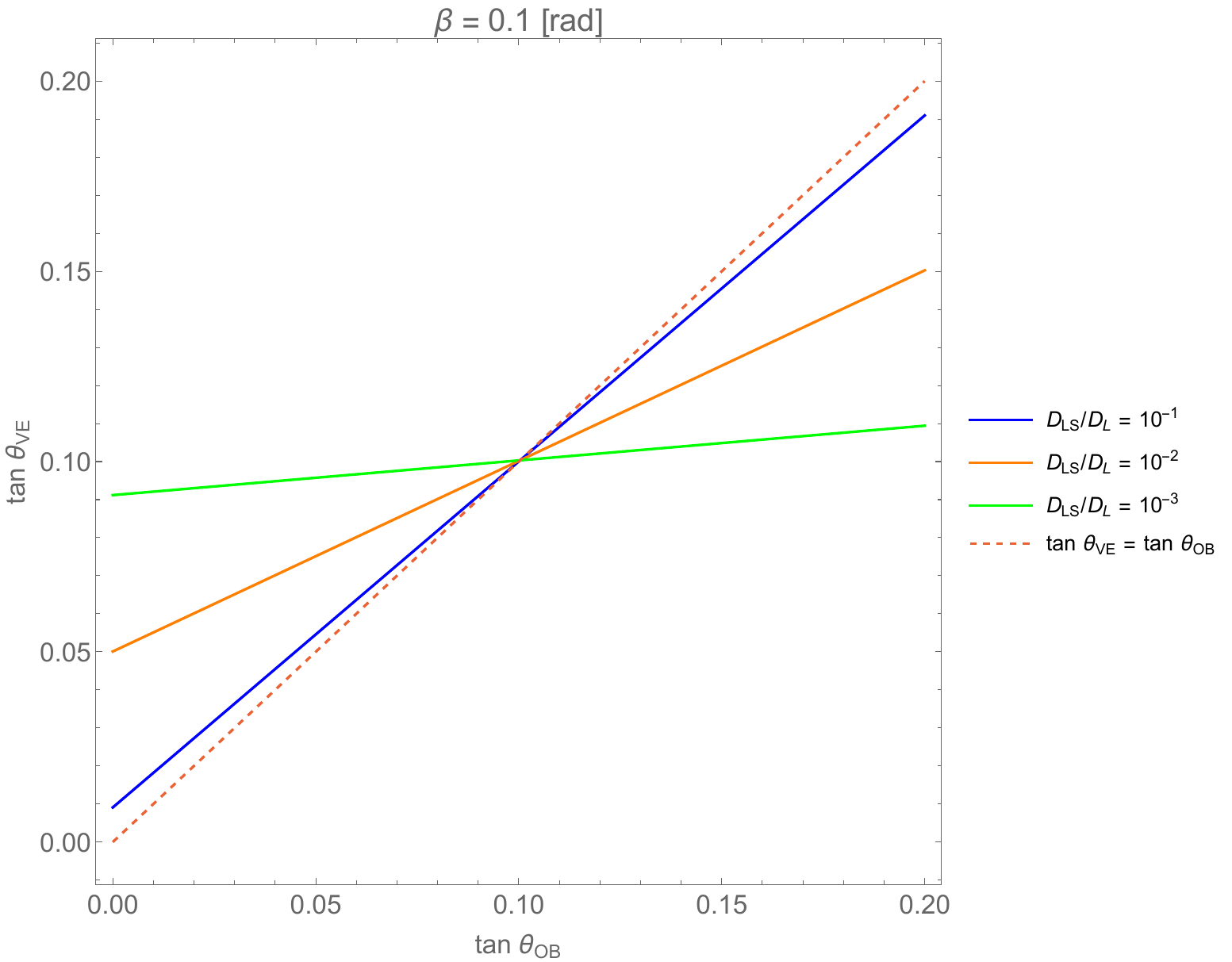}
\caption{
$\tan\theta_{VE}$ and $\tan\theta_{OB}$.
Here, the transformation of $\alpha_{VE}$
is ignored when
we solve the improved VE equation.
The dashed line corresponds to $\tan\theta_{VE} = \tan\theta_{OB}$
as a reference.
$D_{LS}/D_L = 10^{-1}$ (blue), $10^{-2}$ (red), and $10^{-3}$ (green),
where $\beta = 0.1$ is chosen.
}
\label{fig-theta}
\end{figure}

How large are numerical differences
when we ignore the transformation between the two deflection angles?
For this purpose,
we intend to equate $\alpha_{VE}$ and $\alpha_{OB}$.
We focus on the case of $|\tan\alpha_{VE}| \ll1$ and $|\tan\alpha_{OB}| \ll1$.
From Eqs. (\ref{lenseq-VE}) and (\ref{lenseq-Bozza}), then,
we can obtain
\begin{align}
\cfrac{\tan\theta_{VE} - \tan\beta}{\cfrac{D_{LS}}{D_S}(1 + \tan^2\beta)}
=
\cfrac{\tan\theta_{OB} - \tan\beta}{\cfrac{D_{LS}}{D_S} + \tan^2\beta}
+O(\tan^2\alpha_{VE}, \tan^2\alpha_{OB}) ,
\label{difference}
\end{align}
where $\theta_{VE}$ and $\theta_{OB}$ denote
a solution for Eqs. (\ref{lenseq-VE}) and (\ref{lenseq-Bozza}), respectively.
We ignore the second order in $\tan\alpha_{OB}$ below.

For $\tan\theta_{VE}$ to equal to $\tan\theta_{OB}$,
the two denominators in Eq. (\ref{difference}) must be the same as each other.
However, it is not the case.
See Figure \ref{fig-theta} for numerical plots of Eq. (\ref{difference}).

The two denominators are very close to each other,
if and only if
$|\beta| \ll 1$
and $D_L \ll D_{LS}$.
This condition means that
the observer, lens and source are almost coaligned
and the observer is close to the lens.
This condition is not satisfied for most of astronomical situations.

The exceptional case is the gravitational lens by the Sun.
In many astronomical cases ($\beta \neq 0$ and $D_L \gg D_{LS}$),
conversely speaking,
the ignorance of the transformation between the deflection angles
would result in a significant difference between $\tan\theta_{VE}$ and $\tan\theta_{OB}$.

\begin{figure}
\includegraphics[width=8.5cm]{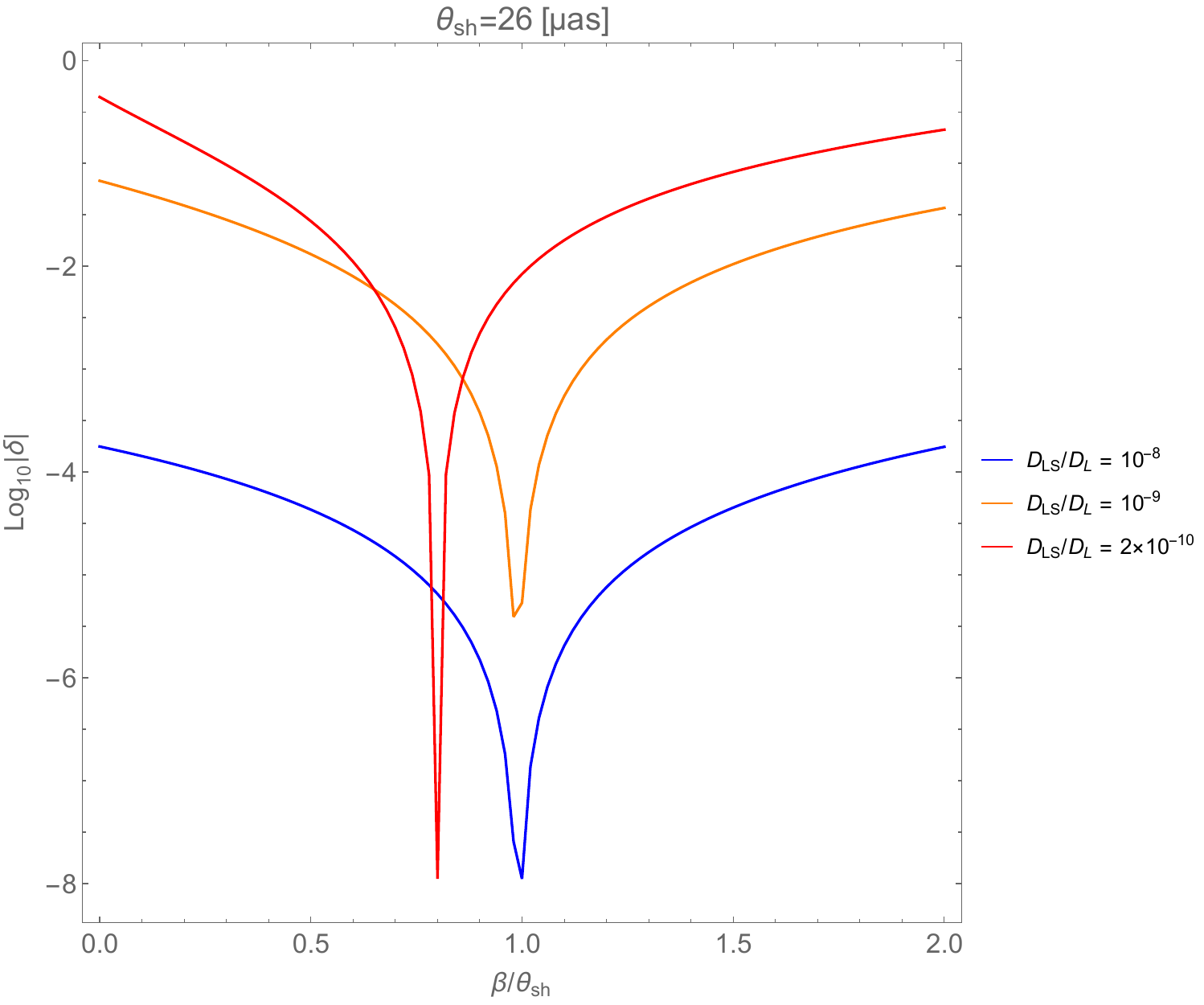}
\caption{
Numerical differences by ignoring the transformation between
$\alpha_{VE}$ and $\alpha_{OB}$ when the improved VE equation is solved.
The parameters correspond to the Sgr A$^*$
\cite{Gillessen}
in the approximation of the Schwarzschild lens,
where the angular radius of the black hole shadow \cite{PT}
is about 26 microarcsec.,
$D_L =8$ kpc,
and
$D_{LS}/D_L = 1.2 \times 10^{-10}$
\cite{EHT2022}.
Here, $\delta \equiv |\theta_{VE}/\theta_{OB} - 1|$ is the relative difference
due to neglecting the transformation.
}
\label{fig-error}
\end{figure}

See Figure \ref{fig-error} for numerical calculations of the difference
when the transformation is ignored.
The numerical plots in this figure are consistent with
an expression as
\begin{align}
\tan\alpha_{VE} - \tan\alpha_{OB}
=
\frac{D_L}{D_{LS}} \sin^2\beta \tan\alpha_{OB}
+ O\left(\tan^2\alpha_{OB}\right) ,
\label{approx}
\end{align}
which can be obtained from Eq. (\ref{transformation})
in the approximation as $|\tan\alpha_{OB}| \ll 1$.
Eq. (\ref{approx}) is consistent with also
Figure \ref{fig-alpha} in the domain of small $\tan\alpha_{OB}$.

\section{Summary}
We have shown that there exists an unphysical branch in VE equation.
VE equation is improved by removing the unphysical branch.
The improved VE equation has been found to be the same as OB equation
when the transformation between the deflection angles is suitably done.
We have argued possible numerical differences
when the transformation is ignored.
In particular,
numerical differences can be significant
in the use of the VE equation for the Sgr A$^*$.

It is left for future to extend the current result to a stationary and axisymmetric lens.

\begin{acknowledgments}
We thank Shwetketu Virbhadra for
helpful comments on the earlier version of our manuscript.
We wish to thank an anonymous referee for helpful comments.
We are grateful to Keita Takizawa and Marcus Werner for fruitful discussions.
We thank Yuuiti Sendouda and Ryuichi Takahashi
for useful conversations.
This work was supported
in part by Japan
Science and Technology Agency (JST) SPRING, Grant
Number, JPMJSP2152 (R.K.)
and in part by Japan Society for the Promotion of Science (JSPS)
Grant-in-Aid for Scientific Research,
No. 20K03963 and 24K07009(H.A.).
\end{acknowledgments}

\appendix
\section{On the {\it minus} branch of VE equation}
This appendix shows that there
exists no solution corresponding to a lensed image
in the minus branch of VE equation
\begin{align}
\tan\theta
=
\frac{D_S (\tan\beta \tan\alpha_{VE} - 1) - \sqrt{\cal D}}{2 D_L \tan\alpha_{VE}}  ,
\label{minusbranch}
\end{align}
in a relativistic compact lens object.
Here, a relativistic compact lens object has a photon sphere
around the lens,
as represented by the Schwarzschild lens.
Therefore, the closest distance $r_0$ of a photon is small, namely
$r_0 \ll D_S$.

VE equation considers the observer and the source in asymptotic flat regions
but not in the neighborhood of the lens.
Hence, $(D_L D_{LS})/(D_S)^2 \ll 1$ is not allowed.
VE equation assumes also a nearly coaligned configuration,
namely a small $\tan\beta$.
Hence, it suffices to assume $|\tan\beta| < 1$ in this appendix.

For the convenience in the following calculations,
we introduce $r \equiv D_L/D_S$,
$s \equiv \tan\beta$ and $t \equiv (\tan\alpha_{VE})^{-1}$
to reexpress
Eq. (\ref{minusbranch}) as
\begin{align}
\tan\theta = \frac{s - f(t)}{2r} ,
\label{minusbranch2}
\end{align}
where $D_{LS} = (1-r) D_S$, and
\begin{align}
f(t) \equiv t + \frac{t}{|t|} \sqrt{\tilde{\cal D}} ,
\label{f}
\end{align}
$\tilde{\cal D} \equiv t^2 {\cal D}/(D_S)^2
= p + q t + t^2$
for $p \equiv s^2 + 4 r (1-r)$ and $q \equiv -2s (1-2r)$.
Note that $0< r <1$, $0 < r(1-r) \leq 1/4$, and $4p > q^2$.
Without loss of generality,
we can choose $s \geq 0$.
We shall examine separately (1) $t > 0$ and (2) $t<0$ below.

\noindent
(1) $t > 0$ case.
Then,
$f(t)
=  t + \sqrt{p + q t + t^2}$,
for which
$df(t)/dt = (\sqrt{p + q t + t^2} + t +q/2)/\sqrt{p + q t + t^2}$.
Here, $\sqrt{p + q t + t^2} + t +q/2 > 0$, because $4p > q^2$.
Hence, $df(t)/dt > 0$ for $t > 0$.
Hence, $f(t) > f(0) = \sqrt{p}$.

We thus obtain
\begin{align}
| \tan\theta |
&=
\frac{1}{2r}
\left| s - f(t) \right|
\notag\\
&= \frac{1}{2r} \left( f(t) - s \right)
\notag\\
& >  \frac{1}{2r} (\sqrt{p} -s) ,
\label{inequality-1}
\end{align}
where $f(t) > \sqrt{p} > s$ is used in the second and third lines.

For $s <1$,
$\sqrt{p} -s > \sqrt{1+4r(1-r)} - 1 = (4 r(1-r))/(1+\sqrt{1 + 4 r(1-r)})
\geq  (4 r(1-r))/(1+\sqrt{2})$,
where $1+4r(1-r) \leq 2$ is used for the last inequality.
By using this inequality for the last line in Eq. (\ref{inequality-1}),
$|\tan\theta| >  (2/(1+\sqrt{2}))(D_{LS}/D_S)$,
which leads to
the closest distance of a photon
$r_0 = |\tan\theta| \times D_L
> (2/(1+\sqrt{2}))(D_L D_{LS}/D_S)$.
However, this inequality contradicts with $r_0 \ll D_S$,
because $(D_L D_{LS})/(D_S)^2 \ll 1$ is not allowed.
For $t > 0$, therefore,
the minus branch
has no solution corresponding to a lensed image.

\noindent
(2) $t < 0$ case.
Then,
$f(t)
=  t - \sqrt{p + q t + t^2}$,
for which
$df(t)/dt = (\sqrt{p + q t + t^2} - t -q/2)/\sqrt{p + q t + t^2}$.
Here, $\sqrt{p + q t + t^2} - t -q/2 > 0$, because $4p > q^2$.
Hence, $df(t)/dt > 0$ for $t < 0$.
Hence, $f(t) < f(0) = -\sqrt{p}$.

We thus find
\begin{align}
| \tan\theta |
&=
\frac{1}{2r}
\left| s - f(t) \right|
\notag\\
&= \frac{1}{2r} (|f(t)| + s)
\notag\\
& >  \frac{1}{2r} (\sqrt{p} +s) ,
\label{inequality-2}
\end{align}
where $f(t) < -\sqrt{p} < 0$ is used in the second and third lines.

By using $\sqrt{p} +s \geq 2\sqrt{r(1-r)}$ for Eq. (\ref{inequality-2}),
we find
$|\tan\theta| >  \sqrt{D_{LS}/D_L}$,
which leads to
the closest distance of a photon
$r_0 = |\tan\theta| \times D_L >  \sqrt{D_L D_{LS}} $.
However, this inequality contradicts with $r_0 \ll D_S$,
because $(D_L D_{LS})/(D_S)^2 \ll 1$ is not allowed.
For $t < 0$ also, therefore, the minus branch has no solution
corresponding to a lensed image.

In either case,
the minus branch
has no solution corresponding to a lensed image.
Therefore, Eq. (\ref{minusbranch}) is unphysical.


\begin{thebibliography}{99}
\bibitem{SEF}
P. Schneider, J. Ehlers, and E. E. Falco,
{\it Gravitational Lenses}
(Springer, NY, 1992).
\bibitem{Petters}
A. O. Petters, H. Levine, and J. Wambsganss,
{\it Singularity Theory and Gravitational Lensing}
(Springer, NY, 2012).
\bibitem{Dodelson}
S. Dodelson,
{\it Gravitational Lensing}
(Cambridge Univ. Press, NY, 2017).
\bibitem{Keeton}
A. B. Congdon, and C. R. Keeton,
{\it Principles of Gravitational Lensing}
(Springer, NY, 2018).
\bibitem{Will}
C. M. Will, Living Rev. Relativity, {\bf 17}, 4 (2014).
\bibitem{EHT2019}
K. Akiyama et al. (Event Horizon Telescope Collaboration),
Astrophys. J. {\bf 875}, L1 (2019);
Astrophys. J. {\bf 875}, L2 (2019);
Astrophys. J. {\bf 875}, L3 (2019);
Astrophys. J. {\bf 875}, L4 (2019);
Astrophys. J. {\bf 875}, L5 (2019);
Astrophys. J. {\bf 875}, L6 (2019).
 \bibitem{EHT2022}
  K. Akiyama et al. (Event Horizon Telescope Collaboration),
  Astrophys. J. {\bf 930}, L12 (2022);
  Astrophys. J. {\bf 930}, L13 (2022);
  Astrophys. J. {\bf 930}, L14 (2022);
  Astrophys. J. {\bf 930}, L15 (2022);
  Astrophys. J. {\bf 930}, L16 (2022);
  Astrophys. J. {\bf 930}, L17 (2022).
\bibitem{Darwin}
C. Darwin, Proc. R. Soc. A {\bf 249}, 180 (1959).\bibitem{Bozza}
V. Bozza,
Phys. Rev. D {\bf 66}, 103001 (2002).
\bibitem{Perlick}
V. Perlick,
Phys. Rev. D {\bf 69}, 064017 (2004).
\bibitem{ngEHT}
https://www.ngeht.org/
\bibitem{VE}
K. S. Virbhadra, and G. F. R. Ellis,
Phys. Rev. D {\bf 62}, 084003 (2000).
\bibitem{Virbhadra2022}
K. S. Virbhadra,
Phys. Rev. D {\bf 106}, 064038 (2022).
\bibitem{Virbhadra2024}
K. S. Virbhadra,
Phys. Rev. D {\bf 109}, 124004 (2024).
\bibitem{Ohanian}
H. C. Ohanian, Am. J. Phys. {\bf 55}, 428 (1987).
\bibitem{Bozza2008}
V. Bozza,
Phys. Rev. D {\bf 78}, 103005 (2008).
\bibitem{GW}
G. W. Gibbons, and M. C. Werner,
Class. Quant. Grav. {\bf 25}, 235009 (2008).
\bibitem{Ishihara2016}
A. Ishihara, Y. Suzuki, T. Ono, T. Kitamura, and H. Asada,
Phys. Rev. D {\bf 94}, 084015 (2016).
\bibitem{Ishihara2017}
A. Ishihara, Y. Suzuki, T. Ono, and H. Asada,
Phys. Rev. D {\bf 95}, 044017 (2017).
\bibitem{Takizawa2020}
K. Takizawa, T. Ono, and H. Asada,
Phys. Rev. D {\bf 102}, 064060 (2020).
\bibitem{DS}
M. P.  Dabrowski, and F. E. Schunck,
Astrophys. J. {\bf 535}, 316 (2000).
\bibitem{Gillessen}
S. Gillessen et al.
Astrophys.J. {\bf 837}, 30 (2017).
\bibitem{PT}
V. Perlick, and O. . Tsupko,
Phys. Rep., {\bf 947}, 1 (2022).
\end{thebibliography}
\end{document}